\documentstyle[preprint,tighten,aps]{revtex} % Specifies the document style.
\begin{document}           %
\draft
\preprint{\vbox{\noindent
To appear in Physics Letters B\hfill hep-ph/9503304\\
          \null\hfill MIT CTP\#2359 \\
          \null\hfill INFNCA-TH-94-23}}
\title{Dimensional Reduction at High Temperature for Fermions}
\author{Suzhou Huang$^{(1,2)}$\cite{email}
        and Marcello Lissia$^{(1,3)}$\cite{email} }
\address{
$^{(1)}$Center for Theoretical Physics, Laboratory for Nuclear Science
and Department of Physics, \\
Massachusetts Institute of Technology, Cambridge, Massachusetts 02139\\
$^{(2)}$Department of Physics, FM-15, University of Washington,
Seattle, Washington 98195~\cite{present}\\
$^{(3)}$Istituto Nazionale di Fisica Nucleare,
via Negri 18, I-09127 Cagliari, Italy~\cite{present}\\
and Dipartimento di Fisica dell'Universit\`a di Cagliari, I-09124 Cagliari,
Italy
         }
\date{September 1994, revised January 1995}
\maketitle                 % Produces the title.
\begin{abstract}
The concept of dimensional reduction in the high temperature regime
is generalized to static Green's functions of composite operators that
contain fermionic fields.
The recognition of a natural kinematic region where the lowest
Matsubara modes are close to their mass-shell, and the ultraviolet
behavior of the running coupling constant of the original theory
are crucial for providing the necessary scale hierarchy.
The general strategy is
illustrated in the asymptotically-free Gross-Neveu model in $1+1$
dimensions, where we verify that dimensional reduction occurs
to the leading order in $g^2(T)$. We also find, in the same model,
that the scale parameter characterizing the dependence on temperature
of the coupling constant in the reduced theory, $\Lambda_T$, is
considerably smaller than $\Lambda_{\overline{\text{MS}}}$.
Implications of our results for QCD are also discussed.
\end{abstract}
\pacs{Keywords: Finite Temperature Field Theory, Dimensional Reduction,
                Scale Parameter}
\narrowtext
\maketitle
\section{Introduction}
\label{intro}
Physics very often simplifies under extreme situations.
For example, specific observables in a $D+1$-dimensional theory
at high temperature ($T$) can sometimes be described by
a $D$-dimensional theory: this phenomenon is known as dimensional
reduction (DR)~\cite{Appelquist81}.
The basic concept of DR is that in the high-$T$ limit all
temporal excitations are naturally ${\cal O}(T)$. If, for either
kinematic or dynamical reasons, there exist modes of order less
than $T$, only these few light modes remain active, while
the others decouple.
This qualitative expectation can be formalized~\cite{Appelquist81} in
some theories, such as QED, order by order
in perturbation theory, analogously to the usual heavy mass
decoupling~\cite{Appelquist75}.

The approach to DR is easier, when there exists a clear scale separation
already at the tree-level. For instance, kinematics makes this scale
hierarchy manifest for observables made by elementary bosonic fields.
The non-zero Matsubara frequencies act like masses of ${\cal O}(T)$
compared to the zero-modes, and these non-zero modes, both bosonic and
fermionic, can be integrated out. If no other dynamical phenomenon occurs,
the result is an effective theory with one less dimension,
which can be used to describe static phenomena of the original theory
in the high-$T$ limit. So far, the existing literature has exclusively
dealt with the dynamics of these bosonic zero
modes~\cite{Nadkarni83,Jourjine84,Alvarez87,Landsman89,Reisz}.

This work will focus on situations where the observables are made
explicitly by fermionic fields, the lowest modes are also
of ${\cal O}(T)$, and, therefore, there is no obvious scale separation.
If at all, the scale separation must be generated dynamically.
We shall systematically study, in the framework of perturbation theory
and renormalization group, how the concept of DR can be generalized
to include these situations. The purpose of deriving an effective
theory for fermions at high-$T$ by integrating out heavy Matsubara modes
is to simplify the physics for screening phenomena involving quarks,
which are well studied in lattice QCD.
Our result provides a formal basis for the
recent interpretation~\cite{Koch92,Hansson92,Schramm93,Koch94,Ishii}
of lattice data
\cite{DeGrand86,DeTar87,Gottlieb87,Koike88,Bitar91,%
Born91,Bernard,Schramm93}
related to screening processes  at high-$T$, which assumes a picture
where only the lowest Matsubara frequencies are important and can be
treated non-relativistically.

After stating the precise criterion for DR for
observables involving fermionic fields, we illustrate the general principle
in the Gross-Neveu model in 1+1 dimensions, and apply it to the
calculation of the screening mass.

In the process, we also calculate the scale parameter that fixes the
dependence of the effective coupling constant on the temperature.
\section{General strategy and criteria for dimensional reduction}
\label{DR}
Let us first recapitulate the criterion for DR when only static
fundamental bosons appear in the external lines.
We say that the $D+1$ dimensional Lagrangian ${\cal L}_{D+1}$
undergoes DR to a specific $D$ dimensional Lagrangian
${\cal L}_D$ as its effective theory, if the following happens.
Static Green's functions of ${\cal L}_{D+1}$ with  small external
momenta ($|\bbox{p}|\ll T$) are equal to the corresponding Green's
functions of ${\cal L}_D$ up to corrections of order
$|\bbox{p}|/T$ and $m/T$, where $m$ is any external dimensionful
parameter in ${\cal L}_{D+1}$, e.g. a mass.
In general, the form and parameters of ${\cal L}_{D}$ are determined
by the original theory.

As stressed by Landsman~\cite{Landsman89},
naive expectations based on tree-level power countings
may fail if there are dynamically generated scales of order $T$.
Nevertheless, these dynamically generated scales must
be proportional to some power of the coupling constant $M\sim g^n T$,
since they are generated by the interaction. Therefore, they
induce corrections of order $M/T\sim g^n$. If $g$ is small, the concept
is still useful, and we say that the reduction is partial.

Situations where composite operators made by fermions appear in the
external lines are of great phenomenological interest: typical
examples are the electromagnetic current, mesonic and baryonic
interpolating fields in QCD. Observables in screening processes
are, for instance, extracted from the spatial correlation functions
of these currents at high $T$ and large distances relative to the
thermal Compton wavelength $1/T$.

When fermions appear in external lines there are two main differences
with the fundamental bosons case. The first difference is that the
lowest Matsubara frequencies for fermions, $\omega_\pm=\pm\pi T$,
are also of order $T$, and hence it is not obvious that they dominate
over the heavier modes. The second difference is that we often
need to consider external momenta of order $T$. In fact, if we want the
fermions to be close to their mass shell in the reduced theory
(this is eventually the physically relevant region), $|\bbox{p}|$
must be of order $T$.

Since $\omega_\pm$ acts in the reduced theory as a large mass,
it has been proposed~\cite{Koch92,Hansson92,Schramm93,Koch94}
that fermions might
undergo a non-relativistic dimensional reduction. This motivates
us to define $\bbox{q}^2\equiv \bbox{p}^2 +(\pi T)^2$, and expand the
Green's functions in the dynamical residual momentum $\bbox{q}^2$ (both
$p_1$ and $q_1$ must be understood in Minkowski space).
In the end, it will be necessary to check the consistency of this
expansion by verifying whether $\bbox{q}^2 \ll (\pi T)^2$. Again,
we expect corrections of order $M/T$ and $q/T$, and talk of
partial reduction if we find that $M$ and/or $|\bbox{q}|$ is
proportional to $g^n T$, with $g$ small.

It should be pointed out that there are in fact other singularities
which could appear in the fermionic Greens functions at high $T$,
corresponding to situations where the external momenta are such
that some of the heavy modes are close to their ``mass-shell''.
However the singularities closest to the origin determine the long
distance behavior of spatial correlations. In fact, once the
free correlation has been separated out, and therefore the trivial
mass contribution to the external momentum has been subtracted, the
interesting dynamical behavior of the remaining correlation is
characterized by $q$, the off-shellness from the lowest
free singularity. This residual-momentum dependence of the
correlation function can be measured in lattice QCD simulations,
since the free contribution to the correlations is known exactly.

In analogy with the heavy mass decoupling theorem, the decoupling of the
heavy modes is manifest only in specific subtraction schemes, such as
the BPHZ scheme. A two-step approach better illustrates the need for
a judicious choice of the counterterms.

Let us consider a graph schematically in the original theory renormalized
in a $T$-independent scheme, e.g. the $\overline{\text{MS}}$ scheme.
We can always split it into light and heavy contributions:
$G^{D+1}(q,T) = G_{L}^{D}(q,T) + G_{H}^{D+1}(q,T)$, where $ G_{L}^{D}$
is the contribution of terms where {\em all\/} loop frequencies have
their smallest value. Since there are no infinite frequency sums
$G^D_L$ is actually $D$ dimensional. Then, we expand $G^{D+1}_{H}$
at $q=0$, and keep terms that are not suppressed by powers of $T$:
$G^{D+1}(q,T) = G_{L}^{D}(q,T) + G_{H}^{D+1}(0,T) + O(q/T)$, where we
have supposed that only one term in the expansion survives. When
DR takes place, the local term $G_{H}^{D+1}(0,T)$ contains contributions
that either can be eliminated by changing the renormalization
prescription, or can be generated in the reduced
theory by a finite number of renormalizable vertices.

The above strategy for the decoupling of the heavy modes
is in fact similar to any other decoupling theorem.
Our particular case differs only in the relevant
kinematic regime where heavy Matsubara modes can possibly decouple,
which happens
when the lightest Matsubara modes are close to their ``mass-shell''.

Because of the $T$-dependent renormalization, the parameters in the
reduced graph (and in the effective Lagrangian that generates such
graph) necessarily depend on $T$. The effective Lagrangian, and the
temperature dependence of its parameters are
uniquely determined by the original theory.
\section{Explicit calculation in the Gross-Neveu model}
\label{model}
Let us illustrate the general idea of last section in a concrete
example of dimensional reduction for static spatial correlations of
currents involving fermionic fields. The model we consider is
the Gross-Neveu model in 1+1 dimensions~\cite{Gross74} described
by the Lagrangian
\begin{equation}
{\cal L} = \bar{\psi}i\gamma\cdot\partial\psi
-\bar{\psi}(\sigma+i\pi\gamma_5)\psi
-{N\over 2g^2}(\sigma^2+\pi^2)\, ,
\end{equation}
where $\sigma$ and $\pi$ are the auxiliary scalar and pseudoscalar
boson fields, respectively. We study this model in the limit
$N\to\infty$ with the coupling constant $g^2$ fixed. In spite of
some obvious limitations, this model shares several qualitative
features with QCD, and it has been often used to test new concepts that
might be relevant for QCD itself.
By Fourier transforming the fields
\begin{equation}
\psi(\tau,{x})={\sqrt{T}}\sum_{n=-\infty}^\infty
\psi_n({x})\/e^{i\omega_n\tau}\, , \quad
\begin{array}{c}
\sigma(\tau,{x}) \\ \pi(\tau,{x}) \end{array}
=\sum_{l=-\infty}^\infty
\begin{array}{c}
\sigma_l({x}) \\ \pi_l({x}) \end{array}
\/e^{i\Omega_l\tau}\, ,
\end{equation}
where $\omega_n=(2n-1)\pi T$ and $\Omega_l=2l\pi T$,
we can rewrite the action as follows~\cite{Jourjine84}
\begin{equation}
\int \!d{x}\!\! \sum_{n,l=-\infty}^\infty \!\Bigg\{
\!\bar{\psi}_n({x})\bigg[\Bigl(-\omega_n\gamma_0
-i{\gamma_1}{\partial_1}\Bigr)\delta_{l0}-\sigma_l({x})
+i\gamma_5\pi_l({x})\bigg]\psi_{n-l}({x})
-{N\delta_{nl}\over 2 g^2T}\bigg[\sigma_l^2({x})+\pi_l^2({x})\bigg]
\Bigg\}.
\label{drl}
\end{equation}
Since we are interested in static Green's functions (zero external
frequency), only terms with $l=0$ are relevant. We are left with a
one-dimensional theory with an infinite number of fermions, each with
a chirally invariant mass $\omega_n$. The tree-level coupling
constant is $g^2 T$. As discussed in the preceding section,
we say that DR occurs, if the static correlations ($l=0$) are reproduced
by the action (\ref{drl}) with only $\omega_n=\pm\pi T$ terms, while
the heavier modes give corrections that either are of
order $1/T$, or can be absorbed by a proper redefinition of the
the parameters of the effective theory, which become dependent
on $T$.

In the symmetric phase (high-$T$) and in the large $N$ limit, this
model has only one non-trivial irreducible graph: the bubble graph.
For static external lines, a calculation along the lines of
Ref.~\cite{Huang93,Huang94} yields
\begin{eqnarray}
i\Pi(p_1)&=&-{NT}\sum_{n=-\infty}^\infty\mu^{2\epsilon}
\int {d^{1-2\epsilon}k_1\over(2\pi)^{1-2\epsilon}}\,{\rm Tr}
\Bigg\{\gamma_5 {i\over k\cdot\gamma}
\gamma_5{i\over(k+p)\cdot\gamma}\Bigg\}\\
&=&-{N\over 2\pi}
\Bigg[{1\over\epsilon}
+\ln\bigg({4\mu^2\, e^{\gamma_E}\over\pi T^2}\bigg)
-{p_1^2\over T^2}\sum_{n=1}^\infty
\frac{1}{(2n-1)^3\pi^2 + (2n-1){p_1^2}/{4T^2}}\Bigg]\, ,
\label{unren}
\end{eqnarray}
where we have used dimensional regularization on the spatial integral.
We are interested in the large distance behavior of the spatial
correlations, which is determined by the lowest singularities, i.e.
$p_1\approx\pm 2\pi T i$. Therefore, we analytically continue
$p_1$ into Minkowski space, and subtract the trivial large-mass
contribution to the external momentum by defining the reduced momentum
$q^2_1=p_1^2 + 4\pi^2 T^2$. Then Eq.~(\ref{unren}) becomes
\begin{eqnarray}
\label{bblr}
\frac{2\pi i}{N}\Pi_{\text{R}}(q^2_1)&=&
-\ln{\left(\frac{\mu^2e^{2\gamma_E}}{\pi^2 T^2}\right)}\nonumber \\
&+&4\left[\left(\frac{q_1}{2\pi T}\right)^2-1\right]
\left[\left(\frac{2\pi T}{q_1}\right)^2 +\sum_{n=1}^\infty
\frac{1}{(2n+1)[4n(n+1)+(q_1/2\pi T)^2]}\right]\, ,
\end{eqnarray}
where the subscript R means that the bubble has been renormalized.
For the sake of concreteness, we used the modified
minimal subtraction ($\overline{\text{MS}}$) scheme, which is
independent of $T$.
Now we assume that $q_1^2\ll T^2$ (this assumption must be verified in
the final results), and expand in $q_1^2/T^2$
\begin{equation}
\frac{2\pi i}{N}\Pi_{\text{R}}(q^2_1)=
-\frac{16\pi^2 T^2}{q_1^2}
-\ln{\left(\frac{\mu^2 e^{2\gamma_E-1}}{16\pi^2 T^2}\right)}
+{\cal O}\left(\frac{q_1^2}{T^2}\right)\, .
\end{equation}

This result illustrates what has been said in the preceding section.
The main momentum dependence $-16\pi^2 T^2/q_1^2$ is given
by the lightest modes, and can be reproduced by a lower dimensional
theory. The other modes give contributions that are
either suppressed by powers of ${q_1^2}/{T^2}$, or momentum
independent, $-\ln{\left({\mu^2 e^{2\gamma_E-1}}/{16\pi^2 T^2}\right)}$.
The large $T$-dependent part of this local piece can be eliminated
by choosing $\mu\propto T$, which corresponds to a $T$-dependent
renormalization scheme.
The specific choice of the proportionality constant
defines the coupling $\tilde{g}^2(T)$, where we use the tilde
to stress the fact that in general $\tilde{g}^2(T)$ is
a different function from $g^2(\mu)$.

The general strategy we adopt to define the temperature dependent
couplings is that the expansion in $q^2/T^2$ of the divergent
one-particle irreducible (1PI) graphs  calculated  at one-loop level
should have no finite corrections in the high-$T$ limit~\cite{Landsman89}.
This is always possible in a renormalizable theory, since we have a
renomalization constant for each divergent 1PI graph (non-divergent
graphs are already suppressed by powers of $q/T$).
The advantages of this definition are that the reduced theory with
these ``optimal'' couplings reproduces exactly all the divergent 1PI
graphs of the original theory to one loop up to power corrections in
$q^2/T^2$, and that it can be easily implemented in any renormalizable
theory, since it only requires perturbative calculations to one loop.
It will become clear later that this choice also minimizes on average
the sub-leading corrections in $\tilde{g}^2(T)$ to physical quantities
in the original theory.

In this particular model the only divergent 1PI graph is the bubble
graph. The requirement that the finite terms vanish implies
$\mu^2=16\pi^2 T^2/e^{2\gamma_E-1}$, which, recalling that
$2\pi/g^2(\mu)\equiv\ln(\mu^2/\Lambda^2_{\overline{\text{MS}}})$,
defines
\begin{equation}
\frac{2\pi}{\tilde{g}^2 (T)}=
\ln{\left(\frac{16\pi^2 T^2}{\Lambda^2_{\overline{\text{MS}}}
                         e^{2\gamma_E-1}}\right) }
\equiv \ln{\left(\frac{T^2}{\Lambda^2_T}\right)} \, ,
\label{gt}
\end{equation}
where we have defined $\Lambda^2_T=
\Lambda^2_{\overline{\text{MS}}}e^{2\gamma_E-1}/(4\pi)^2$. Using the
fact that the critical temperature at which the chiral symmetry gets
restored in this model is
$T_c=\Lambda_{\overline{\text{MS}}}\exp{(\gamma_E)}/\pi$~\cite{Huang94},
$\Lambda_T^2=T^2_c/(16 e)$. It can be
easily checked that the 1-dimensional Lagrangian
\begin{equation}
\int d{x}\, \Bigg\{\sum_{n=\pm}
\bar{\psi}_n({x})\bigg[-\omega_n\gamma_0
-i{\gamma_1}{\partial_1}-\sigma({x})
+i\gamma_5\pi({x})\bigg]\psi_n({x})
-{N\over 2 g^2_1}\bigg[\sigma^2({x})+\pi^2({x})\bigg]
\Bigg\}\, ,
\label{d1l}
\end{equation}
with $g^2_1=T\tilde{g}^2(T)$,
reproduces the bubble graph of the 2-dimensional Lagrangian up to
power-suppressed terms.

Let us now calculate and compare a measurable quantity in both the
original and the reduced theory. The screening mass $\tilde{m}$ is
calculated in both theories by solving the equation
$2\pi/g^2+2\pi i\Pi/N=0$, with  $p_1^2=-\tilde{m}^2$
($q_1^2=(2\pi T)^2-{\tilde m}^2$). In the original theory, by using
Eqs.~(\ref{bblr}) and (\ref{gt}), we find
\begin{equation}
\label{masseq}
\frac{2\pi}{\tilde{g^2}(T)}
-\frac{16\pi^2 T^2}{q_1^2} + 4-\ln{(16)}
+\sum_{n=1}^\infty
\frac{(q_1/2\pi T)^2-1}
{(2n+1)[4n(n+1)+(q_1/2\pi T)^2]}=0\, ,
\end{equation}
whose solution in powers of
$\tilde{g}^2(T)$ is
\begin{equation}
\tilde{m}_{1+1}={2\pi T}\bigg[1-\frac{\tilde{g}^2(T)}{\pi}
+3[\ln{(16)}-3] \left(\frac{\tilde{g}^2(T)}{\pi}\right)^3
+ {\cal O}(\tilde{g}^8(T))\bigg]\, ,
\label{ms2}
\end{equation}
while in the reduced theory we easily find:
\begin{equation}
\tilde{m}_{1}={2\pi T}\bigg[1-\frac{\tilde{g}^2(T)}{\pi}\bigg]\, .
\label{ms1}
\end{equation}
The difference between Eqs.~(\ref{ms2}) and (\ref{ms1}) is a clear
indication that DR is only partial.
This result also explains the physical reason why a partial DR takes place
in this theory, and why this DR implies a non-relativistic reduction.
The screening state is a bound state of a quark and an
antiquark of mass $\pi T$ in one dimension, with a binding energy
$\approx 2\tilde{g}^2(T)T$.
The binding energy in units of the quark mass decreases logarithmically,
therefore our original assumption $q_1^2\ll T^2$ is verified. In
addition,
we expect that the screening mass can be solved from an appropriate
Schr\"{o}dinger equation in the high-$T$ limit. The point here is that
there exists a simplified physical picture, similar to the non-relativistic
reduction assumed by several
authors~\cite{Koch92,Hansson92,Schramm93,Koch94},
because of the fact that
asymptotically free theories possess an additional scale in the large
$T$ limit. This new scale, $T/\ln (T/\Lambda_T)$, makes partial DR
possible. On the contrary, theories with only finite ultraviolet
fixed-points lack this scale hierarchy, and should not undergo even
this partial DR, as we have explicitly verified in the 2+1 Gross-Neveu
model~\cite{long94}.

We point out that our ``optimal'' choice of
$\tilde{g}^2(T)$, i.e. of $\Lambda_T$, makes corrections of order
$\tilde{g}^4(T)$ in Eq.~(\ref{ms2}) disappear. Had we chosen a
different $\Lambda_T$, these corrections would have been there. This
result cannot be universal, since choosing the renormalization scale only
reorganizes the perturbative expansion, and other quantities with
different
expansions cannot be ``fixed'' by a single parameter. Nevertheless,
our perturbative choice of absorbing all finite contributions to the
divergent 1PI graphs in the couplings should at least make the
corrections of order $\tilde{g}^4(T)$ small in most typical quantities.

At last, we draw the attention on the amount of the change of scale:
$\Lambda_T$ is about $0.15 T_c$, which implies that
the perturbative regime sets in much earlier for $T$ than for $\mu$.
This large change of scale can partially be explained by using
$2\pi T$ as the typical temperature energy scale instead of $T$.
This large change of scale is important, since it makes partial
DR useful at relative low temperatures, in spite of
the fact that corrections are suppressed only logarithmically. We
verified that a change of scale of the same magnitude exists in
QCD, and give an explanation of the perturbative behavior of many
quantities already at few~$T_c$~\cite{lambda94}.
\section{Conclusions}
\label{conclu}
We have generalized the concept of dimensional reduction to static
correlations of operators made by fermionic fields (e.g. electromagnetic
current, mesonic currents, etc.).

The basic idea has been illustrated in the Gross-Neveu model, and
applied to the calculation of the screening mass in the high-$T$ limit.

The experience gained in the model study suggests that complete DR
(corrections are suppressed by powers of $T$) for  operators made by
fermionic fields is not possible, since the overall scale is set by $T$.
However, in asymptotically-free theories there appears a new scale,
$\tilde{g}^2(T)T$,  which makes a partial DR possible (corrections are
suppressed by powers of $\tilde{g}^2(T)$).

Nevertheless, this partial DR still yields a simplified physical picture.
The relevant static correlations can be described by the reduced theory,
which, in addition, becomes  non-relativistic in the high-$T$ limit.
This lower dimensional, non-relativistic theory can, for instance,
be used to interpret lattice
data at high-$T$~\cite{Koch92,Hansson92,Schramm93,Koch94}.

We are in the process of explicitly verifying~\cite{QCDdr94} whether
this non-relativistic dimensional reduction also applies to currents
made by fermions in QED and QCD, as the present model
calculation strongly suggests.

We also show that the dependence on the temperature $T$ of
the coupling  constant $\tilde{g}^2(T)$ in the reduced theory is
$\tilde{g}^2(T)=g^2(T\Lambda_{\overline{\text{MS}}}/\Lambda_{T})$, where
$g^2(\mu)$ is the coupling constant in the original theory in the
$\overline{\text{MS}}$ scheme.
An explicit calculation shows that the change of scale
$\Lambda_{\overline{\text{MS}}}/\Lambda_{T}$
is quite large. This result is not typical
of the Gross-Neveu model. In fact we have verified~\cite{lambda94}
that a change of scale of about the same magnitude also appears in QCD.

A rather remarkable consequence of this large reduction of the
scale-parameter going from $\mu$ to $T$ is that
the high-$T$ regime in the reduced theory sets in for temperatures
much lower than naively expected by directly comparing temperatures
to typical perturbative momenta at zero temperature.
For instance, the effective Lagrangian
of finite temperature QCD, which is derived perturbatively
(nevertheless its solution can be non-perturbative),
appears to become reliable at temperatures  as low as about 0.5~GeV
\cite{Reisz,Koch92,Hansson92,Schramm93,Koch94,Ishii}, to be
compared to momenta of about 5~GeV necessary at zero temperature for
perturbation to become reliable.


\begin{thebibliography}{99}
\bibitem[*]{email}
E-mail: shuang@pierre.mit.edu and lissia@pierre.mit.edu

\bibitem[\dag]{present}
Present address.

\bibitem{Appelquist81} T.~Appelquist and R.~D.~Pisarski,
                     Phys. Rev.~D~23 (1981) 2305.

\bibitem{Appelquist75} T.~Appelquist and J.~Carazzone,
                     Phys. Rev.~D~11 (1975) 2856.

\bibitem{Nadkarni83} S.~Nadkarni, Phys. Rev.~D~27 (1983) 917.

\bibitem{Jourjine84} A.~N.~Jourjine, Ann. Phys.~155 (1984) 305.

\bibitem{Alvarez87} Alvarez-Estrada, Phys.~Rev.~D~36 (1987) 2411;
                    Ann. Phys.~174 (1987) 442.

\bibitem{Landsman89} N.~P.~Landsman, Nucl. Phys.~B~322 (1989) 498;
                     Comm. Math. Phys.~125 (1989) 643;
                     E.~L.~M.~Koopman and N.~P.~Landsman, Phys.
                     Lett.~B~223 (1989) 421.

\bibitem{Reisz} T.~Reisz, Z.~Phys.~C~53 (1992) 169;

                P.~Lacock, D.~E.~Miller and T.~Reisz,
                Nucl. Phys.~B~369 (1992) 501.

\bibitem{Koch92} V.~Koch, E.~V.~Shuryak, G.~E.~Brown and A.~D.~Jackson,
                 Phys. Rev.~D~46 (1992) 3169.

\bibitem{Hansson92} T.~H.~Hansson and I.~Zahed, Nucl.
                    Phys.~B~374 (1992) 227.

\bibitem{Schramm93} S.~Schramm and M.~-C.~Chu, Phys. Rev.~D~48 (1993) 2279.

\bibitem{Koch94} V.~Koch, Phys. Rev.~D~49 (1994) 6063.

\bibitem{Ishii} M.~Ishii and T.~Hatsuda, preprint UTHEP-282 (1994).

\bibitem{DeGrand86} T.~A.~DeGrand and C.~E.~DeTar, Phys.
                    Rev.~D~34 (1986) 2469.

\bibitem{DeTar87} C.~E.~DeTar and J.~Kogut, Phys. Rev. Lett.~59 (1987) 399.

\bibitem{Gottlieb87} S.~Gottlieb et al., Phys. Rev. Lett.~59 (1987) 1881.

\bibitem{Koike88} Y.~Koike, M.~Fukugita and A.~Ukawa,
                  Phys. Lett.~B~213 (1988) 497.

\bibitem{Bitar91} K.~M.~Bitar et al., Phys. Rev.~D~43 (1991) 2396.

\bibitem{Born91} K.~D.~Born et al., Phys. Rev. Lett.~67 (1991) 302.

\bibitem{Bernard} C.~Bernard et al., Phys. Rev. Lett.~68 (1992) 2125.

\bibitem{Gross74} D.~Gross and A.~Neveu, Phys. Rev.~D~10 (1974) 3235.

\bibitem{Huang93} S.~Huang, Phys. Rev.~D~47 (1993) 653.

\bibitem{Huang94} S.~Huang and B.~Schreiber, Nucl. Phys.~B426 [FS]
                  (1994) 644.

\bibitem{long94} S.~Huang and M.~Lissia, MIT preprint CTP\#2358 (1994).

\bibitem{lambda94} S.~Huang and M.~Lissia, MIT-CTP\#2360 and
                   hep-ph/9411293, to appear in Nucl. Phys.~B.

\bibitem{QCDdr94} S.~Huang and M.~Lissia, MIT preprint CTP\#2361 (1994).

\end{thebibliography}
\end{document}